\documentclass{article}

\usepackage{microtype}
\usepackage{graphicx}
\usepackage{subcaption}
\usepackage{booktabs} 
\usepackage{multirow}
\usepackage{pifont}
\usepackage{amssymb}
\usepackage{makecell}
\usepackage{booktabs}   
\usepackage{multirow}   
\usepackage{makecell}   
\usepackage{graphicx}   
\usepackage[table]{xcolor} 
\usepackage{pifont}     
\usepackage{array}
\usepackage[table]{xcolor}
\definecolor{hcolor}{HTML}{2E86C1}

\usepackage{hyperref}

\usepackage[preprint]{icml2026}

\newcommand{\tool}[0]{\textsc{TritonDFT}}
\newcommand{\bench}[0]{\textsc{DFTBench}}
\newcommand{\com}[1]{\textcolor{red}{#1}}
\newcommand{\cmark}{\ding{51}}
\newcommand{\xmark}{\ding{55}}

\usepackage{amsmath}
\usepackage{amssymb}
\usepackage{mathtools}
\usepackage{amsthm}
\usepackage[table]{xcolor}
\usepackage[capitalize,noabbrev]{cleveref}
\usepackage{longtable}
\usepackage{booktabs}
\usepackage{array}
\usepackage{caption}
\theoremstyle{plain}

\theoremstyle{definition}

\theoremstyle{remark}

\usepackage[textsize=tiny]{todonotes}
\usepackage{tcolorbox}

\icmltitlerunning{TritonDFT: Automating DFT with a Multi-Agent Framework}

\begin{document}

\twocolumn[
  \icmltitle{TritonDFT: Automating DFT with a Multi-Agent Framework}



  \icmlsetsymbol{equal}{*}

  \begin{icmlauthorlist}
    \icmlauthor{Zhengding Hu}{ucsd}
    \icmlauthor{Kuntal Talit}{ucm}
    \icmlauthor{Zhen Wang}{ucsd}
    \icmlauthor{Haseeb Ahmad}{ucm}
    \icmlauthor{Yichen Lin}{ucsd}
    \icmlauthor{Prabhleen Kaur}{ucsd}
    \icmlauthor{Christopher Lane}{lanl}
    \icmlauthor{Elizabeth A. Peterson}{lanl}
    \icmlauthor{Zhiting Hu}{ucsd}
    \icmlauthor{Elizabeth A. Nowadnick}{ucm}
    \icmlauthor{Yufei Ding}{ucsd}
  \end{icmlauthorlist}

  \icmlaffiliation{ucsd}{University of California, San Diego}
  \icmlaffiliation{ucm}{University of California, Merced}
  \icmlaffiliation{lanl}{Los Alamos National Laboratory}
  
  \icmlcorrespondingauthor{Elizabeth A. Nowadnick}{enowadnick@ucmerced.edu}
  \icmlcorrespondingauthor{Yufei Ding}{yufeiding@ucsd.edu}

  \icmlkeywords{Machine Learning, ICML}

  \vskip 0.3in
]



\printAffiliationsAndNotice{\icmlEqualContribution}

\begin{abstract}

Density Functional Theory (DFT) is a cornerstone of materials science, yet executing DFT in practice requires coordinating a complex, multi-step workflow. Existing tools and LLM-based solutions automate parts of the steps, but lack support for full workflow automation, diverse task adaptation, and accuracy–cost trade-off optimization in DFT configuration. To this end, we present \tool{}, a multi-agent framework that enables efficient and accurate DFT execution through an expert-curated, extensible workflow design, Pareto-aware parameter inference, and multi-source knowledge augmentation. We further introduce \bench{}, a benchmark for evaluating the agent's multi-dimensional capabilities, spanning science expertise, trade-off optimization, HPC knowledge, and cost efficiency.


\tool{} provides an open user interface for real-world usage. Our website is at \url{https://www.tritondft.com}. Our source code and benchmark suite are available at \url{https://github.com/Leo9660/TritonDFT.git}.


\end{abstract}

\section{Introduction}

Density Functional Theory (DFT)~\cite{hohenberg1964dft1, kohn1965dft2} stands as the computational cornerstone of modern materials science. As a first-principles method, DFT provides high-fidelity predictions to validate theoretical hypotheses and reduce experimental cost.


\begin{figure}[t]
    \centering
    \includegraphics[width=0.9\linewidth]{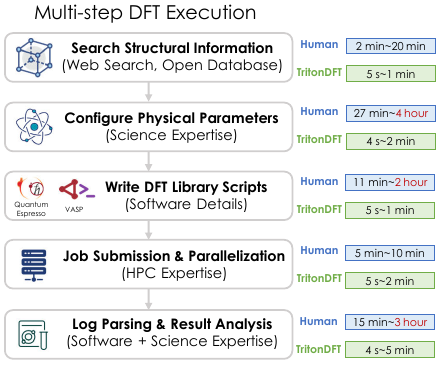}
    \caption{
    DFT execution is a complex, multi-step process requiring heterogeneous domain expertise.
    Based on an internal survey conducted with 19 domain researchers at the PhD level or above, manually handling each step typically takes minutes to hours. \tool{} reduces the per-step time to the scale of seconds to minutes, and provides automation across the entire workflow.
    }
    \label{fig:intro}
\end{figure}

Executing DFT in practice involves a complex, multi-step workflow. Practitioners must search for structural information, configure input parameters, write DFT software-specific scripts, launch and monitor HPC jobs, and interpret and analyze execution results. As shown in Figure~\ref{fig:intro}, such steps require distinct areas of expertise, including physics and  materials science, DFT library software details, and High-Performance-Computing (HPC). Each step takes minutes to hours of manual effort. This imposes substantial overhead and slows down the discovery process. While existing DFT tools can handle certain low-level details, such as script generation~\cite{mathew2017atomate, larsen2017ase} and HPC resource management~\cite{pizzi2016aiida}, users still need to manually handle most of the steps and coordinate the overall workflow.


Such manual overhead gives rise to a natural question: can we leverage Large Language Model (LLM)-based agents to orchestrate these steps and enable automation? LLM agents have been successfully applied across materials science, including domain-specific tool augmentation~\cite{m2024chemcrow, zhang2024honeycomb}, theoretical hypothesis loop~\cite{ding2024matexpert, kumbhar2025hypothesisgeneration}, and autonomous laboratory~\cite{szymanski2023alab, dai2025polybot}. However, applying LLM-based agents to DFT execution, a highly complex, large-scale, and cross-domain theoretical tool, remains challenging and underexplored. These challenges directly motivate our design of \tool{}.

\textbf{First}, the complexity of DFT increases the difficulty to implement a robust and generalizable agent framework. Modern DFT libraries like \texttt{Quantum Espresso}~\cite{giannozzi2009quantumespresso} comprise $>10$ executables and $>50$ commonly used parameters, with completely different invocation patterns and analyzing methods across tasks. While prior work have demonstrated feasibility on specific tasks such as structural relaxation or adsorption~\cite{wang2025dreams, hafner2008vasp}, they typically rely on static, task-specific workflows. In contrast, \tool{} adopts an expert-informed Plan–Execute–Refine workflow design, coupled with an explicit task–to–executable mapping mechanism for extensibility. \tool{} now supports a broad set of tasks, ranging from structural optimization to complex tasks such as phonon properties.




\textbf{Second}, as a numerical simulation, DFT requires results to be obtained both accurately and efficiently. Thus, DFT parameter configurations simultaneously affect numerical fidelity and computational workload, introducing an inherent accuracy–cost trade-off. An existing agent study on parameter configuration~\cite{xia2025agenticDFT} primarily focuses on accuracy while leaving this trade-off problem unsolved. \tool{} introduces a Pareto-aware parameter inference method, which enables the LLM to estimate the accuracy–cost Pareto frontier and iteratively refine configurations. We further incorporate augmentation including domain-specific tools, historical memory mechanism, and an interactive human-in-the-loop interface, to improve the agent’s reliability on the parameter configurations.


\textbf{Third}, despite extensive benchmarks like graduate-level materials-domain knowledge~\cite{zaki2024mascqa, mirza2024superhumanchem}, key capabilities in end-to-end DFT workflows, including numerical accuracy, Pareto-optimality, HPC parallelization, and cost efficiency, remain unevaluated. We present \bench{} to evaluate these capabilities. \bench{} comprises 68 materials spanning 10 distinct types to ensure diversity in physical properties and computational complexity, with expert-curated convergence tests totaling over 500 CPU-hours to obtain
Pareto-optimal parameter configurations and ground-truth calculation results.




In summary, our contribution can be summarized as follows:
\begin{itemize}
    \item We present \tool{}, an expert-informed automation framework for complete DFT workflow execution.
    \item We present \bench{}, a benchmark suite for multi-dimensional capabilities to automate DFT workflows, spanning science, HPC and accuracy-cost tradeoffs. 
    \item Comprehensive experiments demonstrate the potential of \tool{} as a practical building block for real-world research. We also reveal substantial capability differences of LLMs across different task steps.
\end{itemize}

\com{}
\section{Background and Related Work}

\begin{figure}[t]
    \centering
    \includegraphics[width=\linewidth]{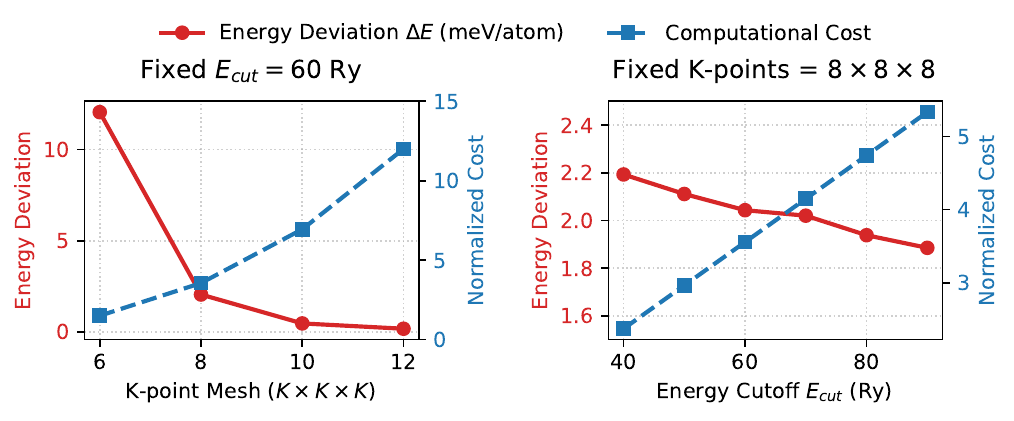}
    
    \caption{Energy Deviation and Computational Cost Variations with different DFT Parameters for Silicon (Space Group $Fd\bar{3}m$). Computational cost is normalized as $C/C_0$, where $C$ denotes the actual execution time and $C_0$ corresponds to the execution time with K=6 and ecut=40.} 
    \label{fig:si_tradeoff}
\end{figure}

\begin{table*}[t]
\centering
\caption{Comparison of \tool{} with state-of-the-art agentic DFT frameworks.}
\label{tab:dft_agent_benchmark_compare}

\small
\renewcommand{\arraystretch}{1.3} 

\scalebox{0.74}{
\begin{tabular}{p{2.6cm}>{\centering\arraybackslash}m{2.6cm}ccccccc}
\toprule

\multirow{2}{*}{\textbf{Method}} &
  \multicolumn{3}{c}{\textbf{Framework Architecture}} &
  \multicolumn{5}{c}{\textbf{Evaluation Dataset \& Metrics}} \\ 
  \cmidrule(lr){2-4} \cmidrule(lr){5-9} 

 &
  \begin{tabular}[c]{@{}c@{}}\textbf{Supported}\\\textbf{Task Types}\end{tabular} &
  \begin{tabular}[c]{@{}c@{}}\textbf{Parameter}\\\textbf{Configuration}\end{tabular} &
  \begin{tabular}[c]{@{}c@{}}\textbf{Knowledge}\\\textbf{Augmentation}\end{tabular} & 
  \begin{tabular}[c]{@{}c@{}}\textbf{Number of}\\\textbf{Material Types}\end{tabular} & 
  \begin{tabular}[c]{@{}c@{}}\textbf{Ground Truth}\\\textbf{Curation}\end{tabular} & 
  \begin{tabular}[c]{@{}c@{}}\textbf{Accuracy-Cost}\\\textbf{Tradeoff}\end{tabular} &
  \begin{tabular}[c]{@{}c@{}}\textbf{Parallel}\\\textbf{Efficiency}\end{tabular} &
  \begin{tabular}[c]{@{}c@{}}\textbf{Monetary}\\\textbf{Cost}\end{tabular} \\ 
\midrule
  
\makecell[l]{DREAMS~\\\cite{wang2025dreams}} &
  \makecell{Surface Chemistry\\(Adsorption)} & 
  \makecell{Physics Only} & 
  \makecell{Open Database} & 
  \makecell{2 (Metal, \\ Insulator)} & 
  \makecell{Public\\Dataset} &
  \xmark &
  \xmark &
  \xmark \\ 
\midrule
  
\makecell[l]{VASPilot~\\\cite{liu2025vaspilot}} &
  \makecell{Electronic Structure\\(Band, DOS)} & 
  \makecell{Physics Only} & 
  \makecell{Open Database} & 
  \makecell{1 (Semiconductor)} & 
  \makecell{Public\\Dataset} &
  \xmark &
  \xmark &
  \xmark \\ 
\midrule
  
\makecell[l]{AgenticDFT~\\\cite{xia2025agenticDFT}} &
  \makecell{Geometry \& Energetics\\(Relaxation, Band)} & 
  \makecell{Physics Only} & 
  \makecell{Open Database} & 
  \makecell{2 (Metal, \\Semiconductor)} & 
  \makecell{Public\\Dataset} &
  \xmark &
  \xmark &
  \xmark \\ 
\midrule
  
\rowcolor{gray!10}
  \textbf{\makecell{\tool{}\\(Ours)}} &
  \textbf{\makecell{General QE Usage\\($> 10$ Types)}} & 
  \textbf{\makecell{Physics + HPC\\(Pareto-aware)}} & 
  \textbf{\makecell{Open Database\\+ Memory\\+ Human Interact.}} & 
  \textbf{\makecell{10 (Metal, Insulator,\\van der Waals,\\ Topological, ...)}} & 
  \textbf{\makecell{Expert Curated\\Calculation}} & 
  \textbf{\cmark} &
  \textbf{\cmark} &
  \textbf{\cmark} \\ 
\bottomrule

\end{tabular}
}
\end{table*}

\subsection{DFT Execution and Challenges}


DFT predicts material properties from first principles, based on the atom identities and structural information. 
Its practical execution entails a multi-step workflow. We formulate a DFT query $\mathcal{Q}$ as a chain of computational steps $\{C_i\}_{i=1}^N$, where each step $C_i = \langle \mathcal{M}_i, \boldsymbol{\theta}_i, \mathcal{O}_i \rangle$ transforms the input state of a specific material via method $\mathcal{M}_i$ (e.g., structure relaxation, self consistent field) into output $\mathcal{O}_i$. $\mathcal{O}_i$ encompasses the updated physical state (e.g., atomic geometry, electron density) and computed observables (e.g., total energy, band gap, density of states). 

The parameter space $\boldsymbol{\theta}_i$ includes two components: \emph{Physical parameters} $\boldsymbol{\theta}_{\mathrm{phy}}$ (e.g., kinetic energy cutoffs, $k$-point densities, and more advanced parameters like spin-orbit coupling) govern numerical accuracy and convergence behavior. These parameters also determine computational workload. \emph{HPC parameters} $\boldsymbol{\theta}_{\mathrm{hpc}}$ (e.g., parallelization over $k$-points or images) determine how parallel tasks are mapped onto hardware resources, influencing execution time and resource efficiency. Thus, efficient DFT execution requires expertise in both materials science and high-performance computing.

\noindent \textbf{Complexity in Manual DFT Execution}. 
Widely used DFT packages such as \texttt{Quantum ESPRESSO}~\cite{giannozzi2009quantumespresso} and \texttt{VASP}~\cite{hafner2008vasp} require substantial manual effort in practice. 
Researchers must retrieve structural data from external sources, construct software-specific input scripts, submit and monitor HPC jobs, and interpret lengthy execution logs. 
Existing DFT tools partially automate this workflow. For example, Atomate~\cite{mathew2017atomate} and ASE~\cite{larsen2017ase} support structure-to-script generation, and AiiDA~\cite{pizzi2016aiida} manages HPC job execution and scheduling. However, existing tools typically operate on parts of the steps and do not provide intelligent guidance for parameter configuration or execution. As a result, DFT researchers still need to manually make decisions and coordinate the overall workflow. 




\noindent \textbf{Parameters with Accuracy--Cost Tradeoff}. 
DFT execution time can range from minutes to hours or even days, which is largely determined by the parameters $\boldsymbol{\theta}$. Figure\com{~\ref{fig:si_tradeoff}} presents one example: increasing some physical parameters like the  $k$-points and energy cutoff reduces calculation error but incurs higher computational cost. 
This introduces a tradeoff: DFT practitioners need to identify \textit{Pareto-optimal} configurations, to maximize the efficiency while meeting the required accuracy. More importantly, the non-trivial coupling among multiple parameters further exacerbates the complexity of this tradeoff. Thus, users must either conduct expensive convergence tests or rely on extensive experience.

\subsection{LLM Agents for Automated Material Discovery}

LLM agents are fundamentally reshaping scientific research by automating complex discovery cycles~\cite{wang2023ai4ssurvey}, enabled by the capabilities of tool-usage~\cite{yao2024tau} and multi-step workflow coordination~\cite{hong2023metagpt}.


\noindent \textbf{Domain-Specific Tool Augmentation}. Prior work augments agents with domain-specific tools, including paper extraction~\cite{cheung2024polyie, hira2024reconstructing, song2023matscinlp}, retrieval augmentation~\cite{schilling2025texttoinsight, chiang2024llamp, mcnaughton2024cactus}, and surrogate models~\cite{liu2024surrogatemodel}. Tool-hub-based agents like HoneyComb~\cite{zhang2024honeycomb} and ChemCrow~\cite{m2024chemcrow} assemble multiple tools to enable coordinated augmentation. These tools typically follow simple input--output protocols and can be effectively encapsulated as single function calls. 
In contrast, using DFT as a tool require more complex workflows, involving physical parameter configuration, HPC job parallelization and management, result analysis and iterative refinement.

\noindent \textbf{Automated Agent Framework}. On the theoretical side, LLM agents have demonstrated promising capabilities in knowledge organization~\cite{tang2025chemagent, ye2024knowledgegraph, shetty2023knowledgesythesis}, property prediction~\cite{song2025predictionandprecursor, yao2025operationalizingDFT}, hypothesis proposal~\cite{ding2024matexpert, kumbhar2025hypothesisgeneration}, and novel material generation~\cite{gruver2024inorganicgeneration, qi2025metascientist, ghafarollahi2025automatingalloydesign, wang2024evolution, jia2024llmatdesign}. On the experimental side, agent-driven embodied systems enable increasingly autonomous wet-lab workflows for synthesis~\cite{szymanski2023alab, delgado2025flowinorganic}, characterization~\cite{dai2025polybot}, and measurement~\cite{olowe2025llmmeasurement, boiko2023coscientist}, substantially expanding the scale of materials discovery~\cite{merchant2023scalingmatdiscovery}.

DFT as a cornerstone should naturally be incorporated into such discovery automation. Recent efforts demonstrate the feasibility of DFT agents~\cite{wang2025dreams, liu2025vaspilot, xia2025agenticDFT}. However, these works are primarily case-study driven, focusing on the correctness on specific tasks and materials, and lack evaluation of accuracy, efficiency and monetary cost. Table~\ref{tab:dft_agent_benchmark_compare} summarizes the key differences between \tool{} and existing DFT agents.

\textbf{Benchmarking Agents in Materials Science}. A growing number of LLM benchmarks has been proposed in material science, including graduate-level question answering~\cite{zaki2024mascqa, mirza2024superhumanchem, zhang2025matscibench, cheung2025msqa}, research-level task completion~\cite{miret2024llmsreadyforrealworlddiscovery, guo2023can8tasks}, specific tool usage~\cite{huang2025cascade}, retrieval-augmented reasoning~\cite{zhong2025ragbenchmarking}, and wet-lab experimental design and analysis~\cite{mandal2025expbench}. However, DFT requires multi-dimensional capabilities including science, HPC, and dual optimization, which are not fully captured by existing benchmarks. This motivates the design of \bench{}.

\begin{figure*}[t]
    \centering
    \includegraphics[width=\linewidth]{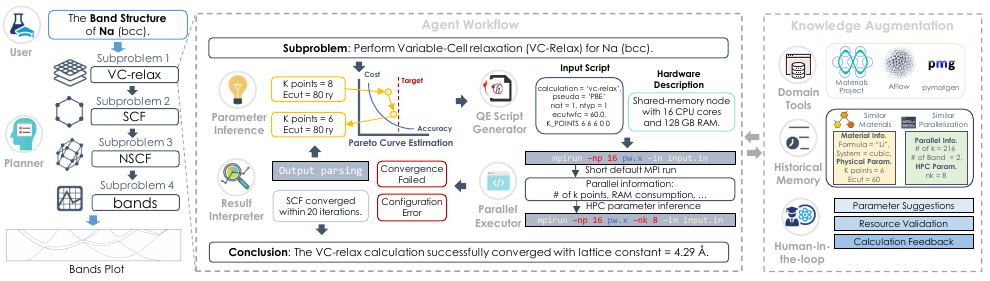}
    \caption{The overview of \tool{} framework.}
    \label{fig:overview}
\end{figure*}

\section{\tool{}: Method and Implementation}

We propose \tool{}, a multi-agent framework designed to achieve fully automated DFT calculations. The framework integrates an LLM-driven agent workflow with \texttt{Quantum Espresso}~\cite{giannozzi2009quantumespresso}, one of the most prominent open-source DFT libraries. Shown in Figure~\ref{fig:overview}, \tool{} accepts task descriptions directly in natural language and autonomously orchestrates the entire DFT workflow, allowing researchers to obtain simulation results without specialized knowledge in computational physics.

\tool{} was developed through a close collaboration between theoretical material scientists and computer science researchers. The agent workflow encapsulates the empirical routines typically employed by experienced physicists. The input prompts are rigorously refined based on both official library documentation and expert domain knowledge.



\subsection{Agentic Workflow for Automation}

We adopt a Planner Agent that dynamically decomposes high-level user queries into a sequence of DFT subproblems. Each category establishes a deterministic mapping to a specific binary executable within the DFT library suite. For instance, the structural relaxation (VC-relax) and self-consistent-field calculation (SCF) are mapped to \texttt{pw.x}. This design ensures every subproblem to be solved through executable invocations. Each executable is registered with structured metadata describing input/output formats and execution constraints. This enables new executables to be added in a modular manner.

Within each subproblem, \tool{} executes a closed-loop solving workflow: (1) \textbf{Parameter Inference} leverages Pareto-optimal-aware reasoning to identify $\boldsymbol{\theta}_{\mathrm{phy}}$ for expected accuracy-cost tradeoff; (2) \textbf{Script Generator} synthesizes these parameters into syntactically correct input files; (3) \textbf{Executor} performs job submission and setup $\boldsymbol{\theta}_{\mathrm{hpc}}$; (4) \textbf{Interpreter} parses the raw output to provides correctness and produces refinement suggestions or summaries.


\noindent \textbf{Pareto-Aware Parameter Refinement}. To navigate the inherent accuracy-cost trade-off and address the bi-objective optimization problem, we introduce a \textit{Pareto-aware Reasoning} mechanism. Instead of static, one-shot inference, the agent is designed to iteratively infer parameter configurations based on result estimation. The agent estimates the discrepancy between the numerical accuracy of the current configuration and user expectations, and closes the feedback loop through iterative parameter self-refinement, ultimately converging toward an estimated Pareto-optimal point. The agent also leverages execution-based feedback by validating DFT outputs against correctness and convergence criteria. Configurations that fail to achieve convergence are treated as invalid points in the accuracy–cost space and excluded from Pareto consideration. Such a refinement process allows the agent to iteratively adjust parameters rather than making a single-shot guess.

\noindent \textbf{Automated Parallelization}. 
The executor automatically sets up parallelization parameters to maximize execution efficiency. It takes as input a natural-language hardware specification (e.g., “32 CPU cores on a 128 GB shared-memory node”), together with the generated input scripts and an estimation of the resource cost to finish the calculation. Following common practices adopted by DFT experts, the agent estimates the resource cost via a short default MPI run (within 30 seconds), and extracts key signals such as number of k-points and total DRAM consumption. These signals provide a lightweight estimation of both parallel scalability and memory footprint, and are jointly used to guide parallelization setup and avoid oversubscription.




\subsection{Domain Knowledge Augmentation}

\tool{} leverages external domain-specific tools and internal memory module for augmented generation.

\noindent \textbf{Domain-Specific Tool Integration}. \tool{} integrates open materials databases, including Materials Project~\cite{jain2013materialsproject} and AFlow~\cite{curtarolo2012aflow}, to retrieve physically grounded information that guides parameter guessing and script construction. The agent dynamically selects specific querying fields based on the current task type. For example, the agent retrieves initial atomic structures for structure relaxation and reasonable lattice constants for subsequent Self Consistent Field. We also integrate \texttt{pymatgen}~\cite{ong2013pymatgen} to enable the agent to perform symmetry analysis and space group verification, ensuring the geometric consistency of the structures.


\noindent \textbf{Historical Memory Mechanism}. Our system implements two forms of memory: For $\boldsymbol{\theta}_{\mathrm{phy}}$, the agents recall parameters from physically similar materials. For successfully converged calculations, the agent summarizes and stores the material information and $\boldsymbol{\theta}_{\mathrm{phy}}$ in memory. During retrieval, the agent first applies structured filtering based on high-level symmetry features, including space group and crystal system. Within the filtered candidates, similarity is computed over other features such as elemental composition, electron count, and unit-cell volume. For $\boldsymbol{\theta}_{\mathrm{hpc}}$, the agent directly compares workload-related features, such as the total number of k-points, the size of the plane-wave basis, and the number of bands. Such historical execution results provides estimation guidance for the accuracy--cost trade-off.

\subsection{User Interface and Interaction}

We adopt a decoupled design that separates the agent backend from the DFT computation backend. This only requires users to connect their own computation platform and specify the task submission method for that environment. Currently, the system supports execution on both local servers and Slurm-managed clusters. 
With this design, \tool{} serves as a comprehensive intermediate layer between the user and the low-level hardware. This relieves users from trivial domain-specific details and tedious manual tasks, such as parameter configuration, parallelization configuration, script writing, progress monitoring, and result analysis.


We have developed an open web interface for \tool{}, allowing users to interact with pure natural language. We also support the human-in-the-loop feedback mechanism: Users can intervene at specific stages to provide feedback, such as reviewing $\boldsymbol{\theta}_{\mathrm{phy}}$ configurations, validating $\boldsymbol{\theta}_{\mathrm{hpc}}$ settings and submission commands, or assessing execution results. This capability enables granular control for researchers with varying levels of expertise, establishing a reliable building block for practical research.





\section{\bench{}: Benchmarking DFT Accuracy and Efficiency with LLM Agents}



\textbf{Benchmark Statistics}. \bench{} is an expert-curated benchmark suite comprising 68 unique crystalline materials, exhibiting diversity in two aspects:



\textit{Physical Diversity}: \bench{} covers 10 distinct material categories, ranging from fundamental electronic phases (Metals, Insulators, Semiconductors) to complex functional and quantum materials (van der Waals, Topological insulators, Ferroelectrics). The material set contains 47 chemical elements, and 23 crystallographic space groups, spanning multiple electronic phases and magnetic ground states. Such diversity in composition, symmetry, and electronic structure
requires the model to understand different material properties and precisely configure $\boldsymbol{\theta}_{\mathrm{phy}}$.



\textit{Computational Complexity Diversity}: The dataset covers a broad range of system sizes, from single-atom primitive cells to polyatomic unit cells, with varying unit cell volumes and total valence electron counts. Such diversity requires the model to efficiently configure $\boldsymbol{\theta}_{\mathrm{hpc}}$ to handle workloads spanning multiple orders of magnitude.


\textbf{Evaluation Design}. Our test suite evaluates the agent in automated DFT workflows along following three dimensions:

\textbf{(i) Accuracy-Cost Trade-off and Pareto-Optimal Parameter Reference }. To evaluate numerical accuracy, we define three target energy deviation levels,
$\Delta E < 1$, 10, and 20 meV/atom. These thresholds reflect commonly adopted DFT accuracy levels under different accuracy--cost trade-offs. The stringent threshold of 1 meV/atom is computational expensive, targeting energy-sensitive properties (e.g., relative phase stability and defect energetics). 10 meV/atom is commonly used for standard high-throughput accuracy (e.g., equilibrium structures and band properties). 20 meV/atom offers the lowest computational cost, commonly used for coarse-grained screening, structure filtering,
and exploratory data generation.


For each material, we provide fully expert-curated input parameter configurations
that satisfy these three accuracy targets, serving as reference Pareto-optimal
configurations. These configurations are obtained through manual convergence testing, involving over 500 CPU-hours of DFT runs to sweep and test relevant
numerical parameters, and identifying optimal configurations on the Pareto curve. We evaluate the agent by comparing its generated parameter configurations
against these reference configurations, assessing its ability to identify
Pareto-optimal points under different energy error tolerances.



\textbf{(ii) HPC Parallelization}. We evaluate the execution time for each test case under a default \texttt{mpirun} setup, where \texttt{np} denotes the total number of parallel processes, and compare it against agent-generated configurations that leverage advanced DFT parallelization parameters such as \texttt{nk} and \texttt{ntg}. This evaluation assesses the agent’s capability of effective parallel acceleration for DFT workloads.


\textbf{(iii) Cost evaluation}. We also measure the number of tokens and the total runtime of LLM calls. These quantities are converted into the corresponding API costs and end-to-end workflow throughput, to evaluate both time and monetary costs of the agent in real-world usage.



\section{Evaluation}

\begin{table*}[t]
\centering
\caption{Model performance on DFT parameter configuration across different LLMs under varying error threshold.}
\label{tab:cost_efficiency}
\definecolor{gc}{gray}{0.85}
\scalebox{0.9}{
\begin{tabular}{lccccccc}
\toprule
\multirow{2}{*}{Model}
& \multicolumn{2}{c}{$\Delta E < $ 20 meV/atom}
& \multicolumn{2}{c}{$\Delta E < $ 10 meV/atom}
& \multicolumn{2}{c}{$\Delta E < $ 1 meV/atom}
& \multirow{2}{*}{\shortstack{Advanced Parameter\\Satisfaction Rate}} \\
\cmidrule(lr){2-3}
\cmidrule(lr){4-5}
\cmidrule(lr){6-7}
& Pass Rate & Norm. Cost
& Pass Rate & Norm. Cost
& Pass Rate & Norm. Cost
& \\
\midrule
GPT 5.2 
& \cellcolor{gc!100}70.5\% & \cellcolor{gc!100}14.29 
& \cellcolor{gc!100}67.0\% & \cellcolor{gc!100}8.95 
& \cellcolor{gc!100}47.1\% & \cellcolor{gc!100}4.23 
& 51.3\%\\

GPT 5.1 
& \cellcolor{gc!56}39.3\% & \cellcolor{gc!44}6.22 
& \cellcolor{gc!49}32.9\% & \cellcolor{gc!47}4.21 
& \cellcolor{gc!21}9.8\% & \cellcolor{gc!53}2.24 
& 43.6\% \\

GPT 4o 
& \cellcolor{gc!75}52.8\% & \cellcolor{gc!13}1.85 
& \cellcolor{gc!57}38.2\% & \cellcolor{gc!14}1.28 
& \cellcolor{gc!29}13.6\% & \cellcolor{gc!12}0.50 
& 28.2\% \\

GPT 4o mini 
& \cellcolor{gc!8}5.7\% & \cellcolor{gc!7}1.01 
& \cellcolor{gc!8}5.6\% & \cellcolor{gc!13}1.17 
& \cellcolor{gc!10}4.5\% & \cellcolor{gc!23}0.97 
& 28.2\% \\

Gemini 2.5 Pro 
& \cellcolor{gc!85}59.6\% & \cellcolor{gc!26}3.77 
& \cellcolor{gc!80}53.9\% & \cellcolor{gc!33}2.95 
& \cellcolor{gc!32}14.9\% & \cellcolor{gc!29}1.24 
& 48.7\% \\

Gemini 2.5 Flash 
& \cellcolor{gc!33}23.6\% & \cellcolor{gc!13}1.85 
& \cellcolor{gc!25}16.9\% & \cellcolor{gc!19}1.68 
& \cellcolor{gc!5}2.3\% & \cellcolor{gc!18}0.78 
& 38.5\% \\

Claude Opus 4.5 
& \cellcolor{gc!13}9.0\% & \cellcolor{gc!11}1.62 
& \cellcolor{gc!8}5.6\% & \cellcolor{gc!15}1.33 
& \cellcolor{gc!10}4.5\% & \cellcolor{gc!14}0.58 
& \textbf{53.8\%} \\

Claude Sonnet 4.5 
& \cellcolor{gc!43}30.3\% & \cellcolor{gc!17}2.38 
& \cellcolor{gc!39}25.8\% & \cellcolor{gc!22}1.93 
& \cellcolor{gc!46}21.6\% & \cellcolor{gc!21}0.87 
& 38.5\% \\
\bottomrule
\end{tabular}
}
\end{table*}

\begin{figure}[t]
    \centering
    \includegraphics[width=\linewidth]{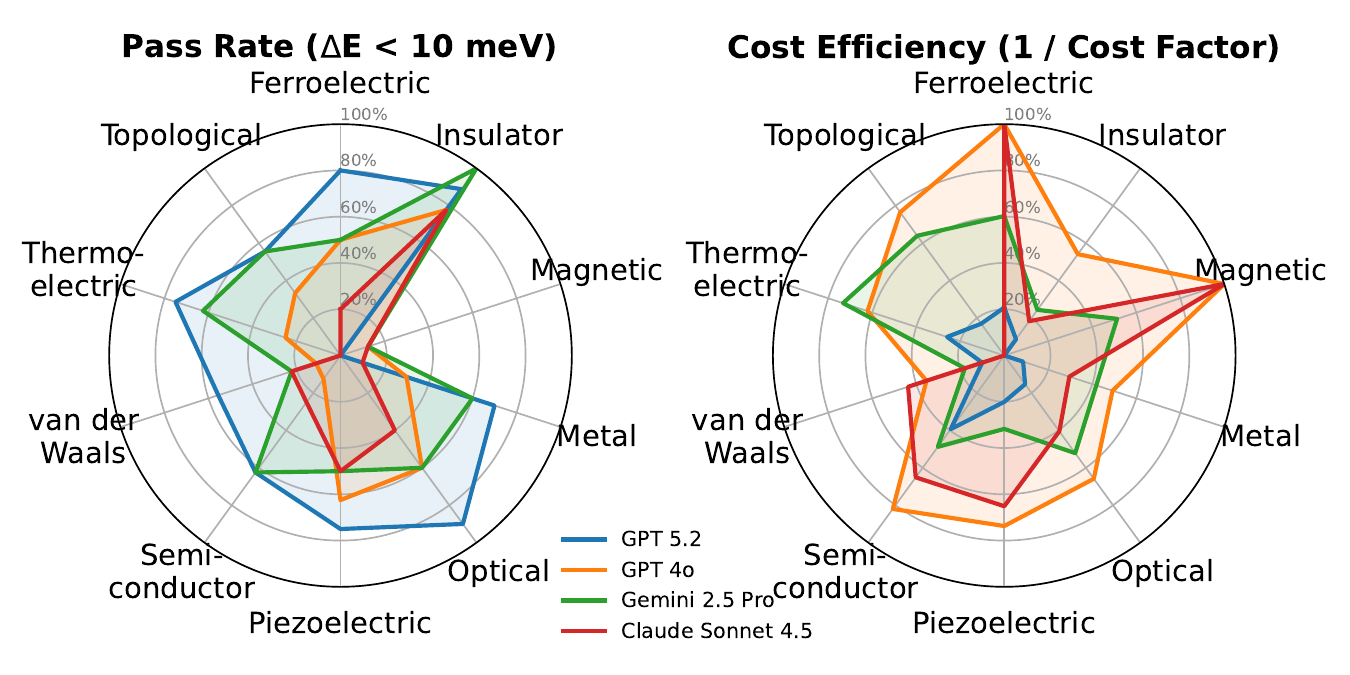}
    \caption{Performance Analysis with Pass Rate and Cost Efficiency across different material types. Cost Efficiency is measured (1 / Cost Factor), averaged over all passed cases within each type.}
    \label{fig:mattypes}
\end{figure}

\subsection{Experimental Setup}

\noindent \textbf{Evaluated Models}. We benchmark eight state-of-the-art models across three major families: OpenAI's GPT 5.2, GPT 5.1, GPT 4o, and GPT 4o mini; Google's Gemini 2.5 Pro and Gemini 2.5 Flash; and Anthropic's Claude Opus 4.5 and Claude Sonnet 4.5. This selection spans a broad spectrum of model architectures and scales, from flagship frontier models to more cost-efficient ones.

\noindent \textbf{Implementation and Platform}. \tool{} integrates \texttt{Quantum ESPRESSO} (v7.4) with a unified API interface for accessing diverse commercial LLMs. The framework is deployed on a high-performance computing node equipped with an AMD EPYC 9534 64-Core Processor clocked at 3.48 GHz. The CPU supports advanced vector extensions like \texttt{AVX-512}, ensuring optimized execution for DFT.

\noindent \textbf{Test Cases}. We perform benchmarking primarily on four DFT tasks: variable-cell relaxation (VC-relax), self-consistent field (SCF), band gap, and density of states (DOS). These tasks cover the most fundamental steps in DFT workflows. We also present examples of more advanced tasks such as phonon analysis in a  case study.

\subsection{Accuracy and Trade-off Analysis}

\noindent \textbf{DFT Parameter Evaluation}. We evaluate the model performance in setting parameters for structural relaxation(VC-relax). It is the most fundamental step that determines the quality of the relaxed geometry and subsequent energy evaluations. Table~\ref{tab:cost_efficiency} summarizes the results. Pass Rate denotes the fraction of cases where the generated configuration satisfies $\Delta E < 20, 10, 1$ meV/atom, compared to the most accurate results obtained with the most stringent configuration obtained by human-performed convergence tests. Normalized Cost is measured as the ratio of computational cost relative to the Pareto-optimal configuration obtained by human-performed convergence tests. Values closer to 1 indicate stronger capability in identifying Pareto-optimal configurations. We also report the Advanced Requirement Satisfaction Rate, corresponding to parameters required by more complex materials such as spin polarization, Hubbard $U$ and van der Waals corrections. In total, 39 materials in \bench{} require one or more such advanced parameters.

\begin{table}[t]
\centering
\caption{Mean relative error (MRE, \%) for VC-relax and SCF, and mean absolute error (MAE) for Band Gap and DOS, computed over successfully finished execution results.}
\label{tab:dft_mae}
\definecolor{gc}{gray}{0.85}
\scalebox{0.9}{
\begin{tabular}{lcccc}
\toprule
Model 
& \makecell{VC-relax\\(Lattice)} 
& \makecell{SCF\\(E/atom)} 
& \makecell{Bandgap\\(eV)} 
& \makecell{E$_F$\\(eV)} \\
\midrule
GPT 5.2
& \cellcolor{gc!15}5.7\% 
& \cellcolor{gc!15}0.0\% 
& \cellcolor{gc!15}0.34 
& \cellcolor{gc!15}1.32 \\
GPT 5.1
& \cellcolor{gc!55}7.4\% 
& \cellcolor{gc!15}0.1\% 
& \cellcolor{gc!15}0.45 
& \cellcolor{gc!55}2.34 \\
GPT 4o
& \cellcolor{gc!100}10.1\% 
& \cellcolor{gc!100}1.2\% 
& \cellcolor{gc!100}1.90 
& \cellcolor{gc!100}8.17 \\
Gemini 2.5 Pro
& \cellcolor{gc!15}5.5\% 
& \cellcolor{gc!15}0.1\% 
& \cellcolor{gc!55}1.02 
& \cellcolor{gc!15}1.45 \\
Gemini 2.5 Flash
& \cellcolor{gc!55}8.1\% 
& \cellcolor{gc!100}0.5\% 
& \cellcolor{gc!55}1.19 
& \cellcolor{gc!100}7.09 \\
Claude Opus 4.5
& \cellcolor{gc!55}8.0\% 
& \cellcolor{gc!55}0.1\% 
& \cellcolor{gc!100}1.84 
& \cellcolor{gc!55}3.01 \\
Claude Sonnet 4.5
& \cellcolor{gc!100}9.1\% 
& \cellcolor{gc!55}0.1\% 
& \cellcolor{gc!100}1.83 
& \cellcolor{gc!55}1.75 \\
\bottomrule
\end{tabular}
}
\end{table}


Among the evaluated models, GPT 5.2 consistently achieves the highest pass rates across all thresholds. This advantage comes at the expense of higher computational cost (with average cost up to 14.29). Intermediate models, including GPT 5.1, GPT 4o, and Gemini 2.5 Pro, exhibit more balanced behavior, with moderate pass rates and lower costs with 20,10 meV/atom. However, under the strict 1 meV/atom threshold, pass rates of these models drop below 15\%. We further observe that the Claude 4.5 family exhibits lower pass rates, with Opus underperforming Sonnet. Based on parameter-level analysis, Opus 4.5 tends to generate aggressively low-cost configurations ($< 1.62$). However, the insufficient accuracy estimation causes suboptimal configuration with frequent threshold violation. 

Regarding advanced parameters, models with stronger reasoning capabilities generally achieve higher satisfaction rates. We also observe that models exhibit distinct strengths across specific parameters. For example, Claude~Opus~4.5 is the most reliable model in Hubbard $U$, while Gemini-2.5-Pro and Gemini-2.5-Flash are the only models that identify cases requiring van der Waals corrections. Interestingly, Opus~4.5 achieves the highest advanced parameter satisfaction rate. This demonstrates its deep domain knowledge, despite weaker cost--accuracy trade-off.



\begin{table}[t]
\centering
\caption{Relative speedup ($\%$) of parallel execution of different LLMs over the default baseline across different CPU core numbers.}
\label{tab:scaling_speedup}
\definecolor{gc}{gray}{0.9} 
\scalebox{0.9}{
\begin{tabular}{lrrc}
\toprule
Model & 16 Cores & 32 Cores & 64 Cores \\
\midrule
GPT 5.2             & \cellcolor{gc!90}+14.4\% & \cellcolor{gc!50}+4.2\% & \cellcolor{gc!94}+15.1\% \\
GPT 5.1             & \cellcolor{gc!70}+11.3\% & \cellcolor{gc!30}-5.8\% & \cellcolor{gc!20}-14.1\% \\
GPT 4o              & \cellcolor{gc!15}-21.0\% & \cellcolor{gc!5}-34.0\% & \cellcolor{gc!12}-23.6\% \\
GPT 4o mini         & \cellcolor{gc!10}-25.7\% & \cellcolor{gc!10}-25.6\% & \cellcolor{gc!45}+2.8\% \\
Gemini 2.5 Pro      & \cellcolor{gc!52}4.74\% & \cellcolor{gc!28}-6.4\% & \cellcolor{gc!35}-3.29\% \\
Gemini 2.5 Flash    & \cellcolor{gc!15}-20.7\% & \cellcolor{gc!2}-43.7\% & \cellcolor{gc!8}-32.0\% \\
Claude 4.5 Opus     & \cellcolor{gc!95}\textbf{+15.4\%} & \cellcolor{gc!100}\textbf{+16.1\%} & \cellcolor{gc!100}\textbf{+16.1\%} \\
Claude 4.5 Sonnet & \cellcolor{gc!80}+13.0\% & \cellcolor{gc!55}+5.1\% & \cellcolor{gc!42}+2.43\% \\
\bottomrule
\end{tabular}
}
\end{table}


\noindent \textbf{Performance across Material Types}. We further compare model performance across different material types in \bench{}, as shown in Figure~\ref{fig:mattypes}. We observe that relatively simple systems, such as metals, semiconductors, and insulators, achieve higher pass rates across models. These materials exhibit weak sensitivity to physical parameters and smoother convergence behavior. In contrast, performance degrades on more complex systems, particularly magnetic materials, where all models exhibit low pass rates ($< 6\%$). Magnetic systems require careful treatment of spin polarization and magnetic ordering, and are highly sensitive to the choice of initial magnetic moments and convergence parameters. We also notice that GPT~5.2 performs better on ferroelectric and optical materials, indicating higher reliability on moderately complex systems.

In terms of cost efficiency, GPT 4o performs best, while Gemini 2.5 Pro strikes a more balanced trade-off between pass rate and cost efficiency across different types. We also observe that for materials such as van der Waals and insulators, cost efficiency tends to be lower. Although these systems are easier to achieve numerical convergence, models still tend to adopt more conservative parameter configurations, leading to unnecessary computational cost.





\noindent \textbf{End-to-End Workflow Evaluation}. Table~\ref{tab:dft_mae}  reports the mean relative error of final results for different task workflows, using the maximum relative deviation of lattice parameters for VC-relax and total energy per atom for SCF. For bandgap calculations and DOS, we report mean absolute error, measured by band gap and Fermi level respectively, since many ground-truth band gaps and Fermi levels are zero, making relative error ill-defined. All results are compared against the most stringent human-configured results. Across all four tasks, stronger models achieve consistently high accuracy, demonstrating the reliability of LLM-based agents for such fundamental tasks in end-to-end DFT workflows.


\subsection{Time and Cost Efficiency Analysis}
\noindent \textbf{Automatic Parallelization}. We evaluate the execution efficiency of the structure relaxation phase, which dominates over 50\% of the end-to-end runtime. Table~\ref{tab:scaling_speedup} shows speedups of LLM-generated parallel configurations over the default MPI baseline (\texttt{mpirun -np <cores>}).

Models with superior coding and reasoning capabilities, such as Claude 4.5 Opus and GPT 5.2, deliver consistent performance gains, achieving peak speedups of up to 16.1\%. This confirms that effective HPC parallelization requires deep reasoning to map physical tasks onto hardware specifications. In contrast, mid-tier and smaller models (e.g., GPT-4o, Gemini 2.5 Flash) exhibit performance degradation. Our analysis reveals that these models attempt aggressive optimizations without fully grasping strict parallel constraints, leading to suboptimal configurations, such as over-parallelization on small-scale systems, or setting a number of k-points \texttt{nk} not divisible by the number of pools.

As the number of cores increases, the complexity of identifying valid parallel parameters rises due to stricter constraints on task divisibility and communication overhead. Most high-performing models exhibit varying degrees of performance degradation at larger scales. Notably, Claude 4.5 Opus stands out as the most robust model, maintaining consistent peak speedups ($\approx$16\%) across all hardware setups.

\begin{figure}[t]
    \centering
    \includegraphics[width=\linewidth]{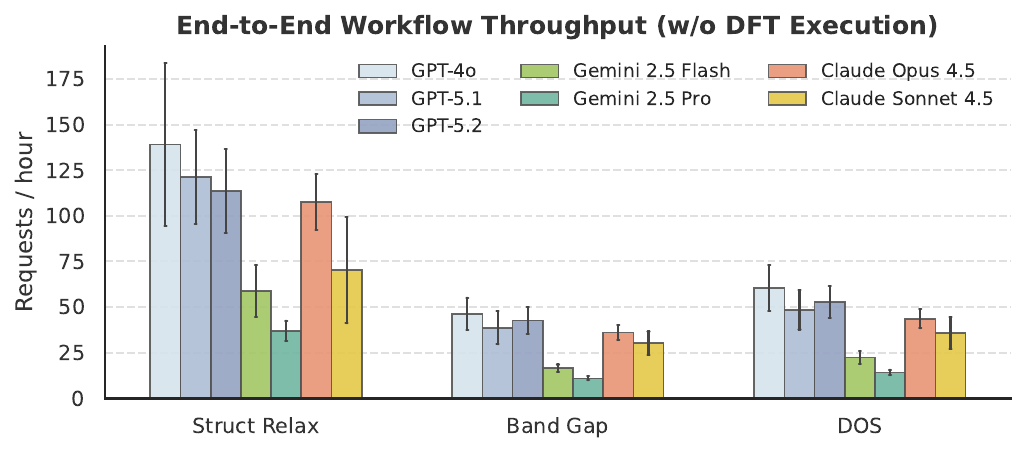}
    \caption{Comparison of \tool{}'s workflow throughput with different LLMs across different DFT tasks.}
    \label{fig:e2e_latency}
\end{figure}

\noindent \textbf{Workflow Throughput.} Figure~\ref{fig:e2e_latency} illustrates the end-to-end throughput of the DFT setup workflow. We exclude the execution time of DFT and focus only on the automation overhead. \tool{} achieves an effective request rate of approximately 10–100 queries/hour, depending on task complexity and API latency. In contrast, our survey on PhD-level practitioners indicates that manually setup DFT workflows sustain less than 1 request/hour. Overall, \tool{} delivers $>10\times$ efficiency improvement.




\noindent \textbf{Monetary Cost}. We calculated the input and output tokens of \tool{} across three tasks and estimated the monetary cost based on official API pricing. The results are shown in Table~\ref{tab:total_cost}. 
Different models exhibit significant divergence in economic efficiency. Gemini 2.5 Flash demonstrates exceptional cost-effectiveness, maintaining an average cost as low as \$0.01--0.03 per query. The GPT-5 series and Gemini 2.5 Pro show comparable pricing with \$0.04--\$0.18 per query. In contrast, the Claude series incurs the highest costs, reaching up to \$0.44 per query for complex calculations.




\begin{table}[t]
\centering
\caption{Average cost consumption (USD) per query regarding API usage across different tasks.}
\label{tab:total_cost}
\definecolor{gc}{gray}{0.9}
\scalebox{0.85}{
\begin{tabular}{lccc}
\toprule
Model & Struct Relax & Band Gap & DOS \\
\midrule
GPT 5.2           & \cellcolor{gc!33}0.05 $\pm$ 0.02 & \cellcolor{gc!34}0.15 $\pm$ 0.04 & \cellcolor{gc!35}0.13 $\pm$ 0.04 \\
GPT 5.1           & \cellcolor{gc!27}0.04 $\pm$ 0.02 & \cellcolor{gc!30}0.13 $\pm$ 0.04 & \cellcolor{gc!27}0.10 $\pm$ 0.03 \\
GPT 4o            & \cellcolor{gc!40}0.06 $\pm$ 0.03 & \cellcolor{gc!41}0.18 $\pm$ 0.04 & \cellcolor{gc!38}0.14 $\pm$ 0.03 \\
Gemini 2.5 Pro    & \cellcolor{gc!33}0.05 $\pm$ 0.03 & \cellcolor{gc!30}0.13 $\pm$ 0.04 & \cellcolor{gc!30}0.11 $\pm$ 0.04 \\
Gemini 2.5 Flash  & \cellcolor{gc!7}0.01 $\pm$ 0.01 & \cellcolor{gc!7}0.03 $\pm$ 0.01 & \cellcolor{gc!8}0.03 $\pm$ 0.01 \\
Claude Opus 4.5   & \cellcolor{gc!100}0.15 $\pm$ 0.08 & \cellcolor{gc!100}0.44 $\pm$ 0.13 & \cellcolor{gc!100}0.37 $\pm$ 0.11 \\
Claude Sonnet 4.5 & \cellcolor{gc!100}0.15 $\pm$ 0.06 & \cellcolor{gc!77}0.34 $\pm$ 0.09 & \cellcolor{gc!76}0.28 $\pm$ 0.08 \\
\bottomrule
\end{tabular}
}
\end{table}



\subsection{Case Study}

\begin{figure}[t]
    \centering
    \includegraphics[width=\linewidth]{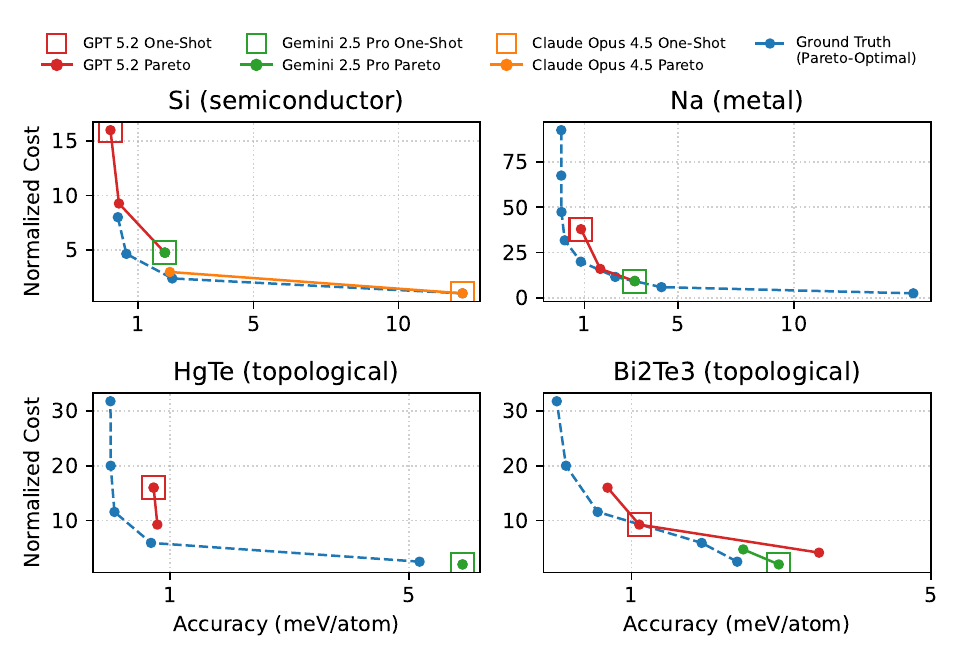}
    \caption{Accuracy-Cost Trade-offs across different models and parameter inference methods when using One-Shot inference and Pareto-aware inference, respectively. } 
    \label{fig:ablation1}
\end{figure}

\noindent \textbf{Pareto-aware Reasoning}. Figure~\ref{fig:ablation1} compares the accuracy--cost trade-offs obtained on representative materials under three target thresholds (20, 10, and 1~meV/atom). We contrast one-shot parameter inference with Pareto-aware iterative reasoning (Pareto). Overall, Pareto-aware reasoning enables models to adaptively navigate the accuracy--cost frontier. GPT~5.2 benefits the most, achieving up to 4.1$\times$ reductions in normalized cost while still meeting the target accuracy constraints. Gemini~2.5~Pro tends to select a fixed, relatively low-accuracy configuration, yet its choices closely follow the ground-truth frontier. In contrast, Claude Opus 4.5 tends to select suboptimal configurations on complex materials (e.g., topological systems), leading to frequent violations of the prescribed accuracy budgets.

\begin{figure}
    \centering
    \includegraphics[width=\linewidth]{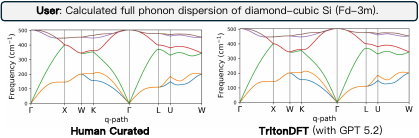}
    \caption{Comparison of \tool{}'s results with human-curated results on phonon dispersion calculation.}
    \label{fig:casestudy}
\end{figure}

\noindent \textbf{Broader Task Types}. \tool{} is designed to handle a broad set of DFT task types. Figure~\ref{fig:casestudy} shows a representative example, where all intermediate steps are automatically handled by the agent except for figure plotting. With our Plan--Execute--Refine workflow, the agent autonomously plans the entire solution path (\texttt{pw.x}--\texttt{ph.x}--\texttt{q2r.x}--\texttt{matdyn.x} in this case) without user specification. This design eliminates the need for defining static and task-specific workflows in prior work~\cite{wang2025dreams}.

We also observe that the agent exhibits clear adaptivity to different user intents by adjusting key numerical parameters. For example, in phonon calculations, the agent dynamically adapts the phonon $q$-point sampling density when performing $\Gamma$-point and full phonon dispersion calculations. In elastic energy--strain analysis that is highly sensitive to numerical noise, the agent employs stricter electronic and force convergence thresholds compared to relaxation tasks.

\section{Conclusion}

In this work, we presented \tool{}, a multi-agent framework that automates Density Functional Theory (DFT) workflows through expert-informed planning, Pareto-aware parameter inference, and automated HPC orchestration. To assess multi-dimensional agentic capabilities in this domain, we introduced \bench{}, a benchmark spanning diverse materials to evaluate numerical accuracy, parallelization efficiency, and cost-effectiveness. Our evaluation confirms that \tool{} delivers a $>$10$\times$ acceleration over manual expert execution. We further identify distinct strengths across models, such as GPT-5.2 excels in accuracy, Gemini~2.5~Flash provides a better accuracy--cost tradeoff, and Opus~4.5 performs better in parallelization schemes.


\noindent \textbf{Future Direction}. Future development will address the observed limitations in modeling complex quantum states, such as magnetic materials, where current models achieve pass rates below 6\%, by integrating specialized physics-informed reasoning modules. We aim to expand the framework's modularity beyond \texttt{Quantum Espresso} to support diverse DFT solvers, enhancing portability across the materials science ecosystem. Additionally, we plan to integrate \tool{} with autonomous experimental platforms for automated theoretical simulation and validation, to enable a low-cost, high-throughput in closed-loop discovery.




\section*{Impact Statement}

This paper presents work whose primary goal is to accelerate scientific discovery by automating complex Density Functional Theory (DFT) workflows. By democratizing access to high-fidelity material simulations, \tool{} has the potential to expedite the development of critical technologies, such as clean and low-energy materials and novel semiconductors. A key societal benefit of this work lies in its emphasis on computational efficiency; the framework’s Pareto-aware parameter inference and automated parallelization are specifically designed to optimize High-Performance Computing (HPC) resource usage, thereby reducing the energy footprint associated with large-scale scientific simulations. 

We explicitly incorporate a human-in-the-loop mechanism to ensure expert oversight and validation of results for two reasons. First, the automation of material design may carry theoretical risks regarding dual-use applications (e.g., the design of hazardous materials). Second, while this automation democratizes access to complex simulation tools for non-experts, the reliance on probabilistic models necessitates robust verification mechanisms to prevent the propagation of scientific errors. We do not foresee other specific negative societal consequences beyond those broadly associated with the application of ML to scientific automation.

\bibliography{ref}
\bibliographystyle{icml2026}

\newpage
\appendix
\onecolumn



\section{Details of \bench{}}

\subsection{Information of Materials}

\bench{} comprises 68 distinct materials spanning 10 different material categories. All material information is summarized in Table~\ref{tab:benchmark_data}.

\footnotesize
\begin{longtable}{c p{1.4cm} p{4.8cm} p{1.4cm} c}
\caption{Benchmark dataset of materials used for DFT workflow evaluation, including material category (multiple categories separated by commas), space group, and number of atoms in the conventional unit cell.}
\label{tab:benchmark_data} \\

\toprule
\# & Material & Category & Space Group & Atoms \\
\midrule
\endfirsthead

\toprule
\# & Material & Category & Space Group & Atoms \\
\midrule
\endhead

\midrule
\multicolumn{5}{r}{Continued on next page} \\
\midrule
\endfoot

\bottomrule
\endlastfoot

1 & CaO & Insulator & Fm$\bar{3}$m & 2 \\
2 & KCl & Insulator & Fm$\bar{3}$m & 2 \\
3 & LiF & Insulator & Fm$\bar{3}$m & 2 \\
4 & MgF$_2$ & Insulator & P4$_2$/mnm & 6 \\
5 & MgO & Insulator & Fm$\bar{3}$m & 2 \\
6 & NaCl & Insulator & Fm$\bar{3}$m & 2 \\
7 & NaF & Insulator & Fm$\bar{3}$m & 2 \\
8 & SiO$_2$ & Insulator & P3$_1$21 & 9 \\
9 & BN (hex) & Insulator & P6$_3$/mmc & 4 \\
10 & C$_{diamond}$ & Insulator & Fd$\bar{3}$m & 2 \\
11 & BN & Insulator, Piezoelectric & Fm$\bar{3}$m & 2 \\
12 & AlN & Insulator, piezoelectric & P6$_3$mc & 4 \\
13 & BeO & Insulator, piezoelectric & P6$_3$mc & 4 \\
14 & BaTiO$_3$ & Insulator, Ferroelectric & P4mm & 5 \\
15 & HfO$_2$ & Insulator, Ferroelectric & Pca2$_1$ & 12 \\
16 & NaNbO$_3$ & Insulator, Ferroelectric & Pm$\bar{3}$m & 5 \\
17 & LiTaO$_3$ & Insulator, Ferroelectric, Piezoelectric & R3c & 10 \\
18 & PbTiO$_3$ & Insulator, Ferroelectric, Piezoelectric & P4mm & 5 \\
19 & Cr$_2$O$_3$ & Insulator, Magnetic & R3c & 10 \\
20 & FeO & Insulator, Magnetic & P4/mmm & 8 \\
21 & MnO & Insulator, Magnetic & P4/mmm & 8 \\
22 & NiO & Insulator, Magnetic & P4/mmm & 8 \\
23 & Al$_2$O$_3$ & Insulator, Optical & R$\bar{3}$c & 10 \\
24 & BaF$_2$ & Insulator, Optical & Fm$\bar{3}$m & 3 \\
25 & CaF$_2$ & Insulator, Optical & Fm$\bar{3}$m & 3 \\
26 & Bi$_2$Se$_3$ & Insulator, van der Waals, Topological & R$\bar{3}$m & 5 \\
27 & Ag & Metal & Fm$\bar{3}$m & 1 \\
28 & Al & Metal & Fm$\bar{3}$m & 1 \\
29 & Au & Metal & Fm$\bar{3}$m & 1 \\
30 & Cu & Metal & Fm$\bar{3}$m & 1 \\
31 & Fe & Metal & I4/mmm & 1 \\
32 & K & Metal & Im$\bar{3}$m & 1 \\
33 & Li & Metal & Im$\bar{3}$m & 1 \\
34 & Mo & Metal & Im$\bar{3}$m & 1 \\
35 & Na & Metal & Im$\bar{3}$m & 1 \\
36 & W & Metal & Im$\bar{3}$m & 1 \\
37 & MgB$_2$ & Metal & P6/mmm & 3 \\
38 & Nb & Metal & Im$\bar{3}$m & 1 \\
39 & Pb & Metal & Fm$\bar{3}$m & 1 \\
40 & Sn ($\beta$-Sn) & Metal & I4$_1$/amd & 2 \\
41 & V & Metal & Im$\bar{3}$m & 1 \\
42 & Co & Metal, Magnetic & P6$_3$/mmc & 2 \\
43 & Fe (bcc) & Metal, Magnetic & Im$\bar{3}$m & 1 \\
44 & Fe$_3$O$_4$ & Metal, Magnetic & Fd$\bar{3}$m & 14 \\
45 & Ni & Metal, Magnetic & Fm$\bar{3}$m & 1 \\
46 & FeSe & Metal, van der Waals & P4/nmm & 4 \\
47 & NbSe$_2$ & Metal, van der Waals & P6$_3$/mmc & 6 \\
48 & AlP & Semiconductor & F$\bar{4}$3m & 2 \\
49 & CdTe & Semiconductor & F$\bar{4}$3m & 2 \\
50 & GaAs & Semiconductor & F$\bar{4}$3m & 2 \\
51 & Ge & Semiconductor & Fd$\bar{3}$m & 2 \\
52 & InP & Semiconductor & F$\bar{4}$3m & 2 \\
53 & Si & Semiconductor & Fd$\bar{3}$m & 2 \\
54 & ZnS & Semiconductor & F$\bar{4}$3m & 2 \\
55 & ZnSe & Semiconductor & F$\bar{4}$3m & 2 \\
56 & PbTe & Semiconductor, Topological & Fm$\bar{3}$m & 2 \\
57 & SnTe & Semiconductor, Topological & Fm$\bar{3}$m & 2 \\
58 & Mg$_2$Ge & Semiconductor, Thermoelectric & Fm$\bar{3}$m & 3 \\
59 & Mg$_2$Si & Semiconductor, Thermoelectric & Fm$\bar{3}$m & 3 \\
60 & SiGe & Semiconductor, Thermoelectric & F$\bar{4}$3m & 2 \\
61 & GeTe & Semiconductor, Ferroelectric & R3m & 2 \\
62 & CdS & Semiconductor, Piezoelectric & P6$_3$mc & 4 \\
63 & CdSe & Semiconductor, Piezoelectric & P6$_3$mc & 4 \\
64 & GaN & Semiconductor, Piezoelectric & P6$_3$mc & 4 \\
65 & ZnO & Semiconductor, Piezoelectric & P6$_3$mc & 4 \\
66 & Bi & Topological & R$\bar{3}$m & 2 \\
67 & Na$_3$Bi & Topological & P6$_3$/mmc & 8 \\
68 & Sb & Topological & R$\bar{3}$m & 2 \\

\end{longtable}
\normalsize

\section{Details of \bench{}}

\subsection{Information of Materials}

\bench{} comprises 68 distinct materials spanning 10 different material categories. 
Some materials belong to multiple categories (e.g., semiconductor, optical). 
All material information is summarized in Table~\ref{tab:benchmark_data}.

\subsection{Question Templates}

\bench{} provides question templates for a diverse set of tasks. 
For each material, the placeholders \{formula\} and \{space group\} in the templates are instantiated accordingly. The required accuracy is also parameterized as \{dEnergy\}, which specifies the target energy tolerance for the task. The question templates used for evaluation are listed below:

\noindent \textbf{VC-relax}: For \{formula\} with space group \{space group\}, perform a variable-cell relaxation (vc-relax) calculation. Require an energy convergence of \{dEnergy\}. Use a PBE functional as pseudopotentials. Return the fully relaxed structure (atomic parameters).

\noindent \textbf{SCF}: For \{formula\} with space group \{space group\}, run a self-consistent field (SCF) calculation and report the converged total energy per formula unit in eV. Require an energy convergence of \{dEnergy\}. Use a PBE functional as pseudopotentials.

\noindent \textbf{Band Gap}: For \{formula\} with space group \{space group\}, Use a full workflow consisting of: 1) vc-relax, 2) scf, 3) nscf, 4) band structure calculation and 5) band post-processing. Require an energy convergence of \{dEnergy\}. Use a PBE functional as pseudopotentials. Do not need to explicitly return band structure. Just finish the calculation.

\noindent \textbf{DOS}: For \{formula\} with space group \{space group\}, use a full workflow consisting of: 1) vc-relax, 2) scf, 3) nscf, 4) band structure calculation 5) band post-processing, and 6) DOS post-processing. Require an energy convergence of \{dEnergy\}. Use a PBE functional as pseudopotentials. Do not need to explicitly return band structure and DOS. Just finish the calculation.

Currently, our evaluation covers the four most essential DFT task types. In the future, we will support additional question templates and datasets. 
Notably, new tasks can be incorporated in a modular manner, making such extensions fully feasible.

\section{Computational details for human-performed DFT calculations}
All density functional theory (DFT) calculations were performed using \textsc{Quantum ESPRESSO}. Structural relaxations were carried out with a total energy convergence threshold of $1.0 \times 10^{-9}$~Ry and a force convergence threshold of $1.0 \times 10^{-8}$~Ry/Bohr. For the self-consistent field (SCF) cycles, the electronic convergence threshold was set to $1.0 \times 10^{-10}$~Ry, and a mixing parameter $\beta = 0.7$ was employed to ensure stable convergence. The plane-wave kinetic energy cutoff and Monkhorst--Pack $k$-point grids were determined from systematic convergence tests, requiring the total energy to be converged within 1~meV per atom. The converged cutoff energy and $k$-point mesh were subsequently used for all subsequent SCF calculations. For non-self-consistent field (NSCF) calculations, denser $k$-point meshes were adopted to obtain accurate electronic properties such as band structures and density of states. The $k$-point density for NSCF calculations was increased on a case-by-case basis depending on the specific property being evaluated. Complete convergence data, including cutoff energies, $k$-point grids, and total energies, are provided in the accompanying CSV file to ensure reproducibility. 

\subsection{Estimation of Normalized cost}

The number of plane waves in a DFT calculation is controlled by the kinetic energy cutoff ($E_{\mathrm{cut}}$), which defines the radius of the reciprocal-space sphere used in the plane-wave expansion. The number of plane waves scales as
\begin{equation}
N_{\mathrm{PW}} \propto E_{\mathrm{cut}}^{3/2},
\end{equation}
while the computational cost scales linearly with the total number of $k$-points, $N_k = k_x k_y k_z$. 

Accordingly, the overall computational cost scales approximately as
\begin{equation}
C \propto E_{\mathrm{cut}}^{3/2} \, N_k.
\end{equation}

To compare different parameter choices, we define a normalized cost metric as
\begin{equation}
C_{\mathrm{norm}} =
\frac{E_{\mathrm{cut}}^{3/2} \, (k_x k_y k_z)_{\mathrm{agent}}}
{E_{\mathrm{cut}}^{3/2} \, (k_x k_y k_z)_{\mathrm{ground\,truth}}}.
\end{equation}

\subsection{Calculation of Band Gap}
Band gaps were computed from the highest occupied molecular orbital (HOMO) and lowest unoccupied molecular orbital (LUMO) energies obtained from the NSCF calculations. For certain materials, electronic smearing was required to achieve stable convergence. In such cases, \textsc{Quantum ESPRESSO} does not explicitly report the HOMO and LUMO energies and instead provides only the Fermi energy. As a result, the band gaps extracted for these systems may appear as zero in Table~\ref{tab:benchmark_data}. These materials require additional post-processing or alternative workflows for an accurate determination of the band gap. 

In this work, our primary objective was to benchmark the agent's performance within a standard automated workflow. Implementation of specialized procedures for robust band gap extraction in smeared calculations is planned for future versions of the framework.

\section{Details about \tool{} Framework Design}

\subsection{QE-aware Multi-step Planner}

The planner currently incorporates 13 executable files from Quantum ESPRESSO. The prompt is designed as follows:

\begin{tcolorbox}[colback=gray!5,colframe=black!50,boxrule=0.5pt,arc=2pt]
\scriptsize

You are a strict planning assistant for Quantum ESPRESSO.

\textbf{Output requirements:}

- Decompose the user query into 1..N subproblems.

- Each subproblem must be wrapped as 
\texttt{\textless subproblem1\textgreater...\textless/subproblem1\textgreater}, 
\texttt{\textless subproblem2\textgreater...\textless/subproblem2\textgreater}, etc. (in order).

- Each subproblem must contain four fields:

Problem: What to calculate, Tool: Tool to use, Required input: Required input parameters (Do not give any concrete parameter value here, just describe what is needed). These fields MUST appear on separate lines, each separated by a newline; otherwise, the output is considered incorrect.

- Keep each subproblem short (2--3 lines).

- Do not output anything outside 
\texttt{\textless subproblem\textgreater} blocks.

\textbf{Core rules:}

Allowed tools: \texttt{pw\_scf, pw\_nscf, pw\_relax, pw\_vc\_relax, pw\_bands, bands\_post, dos\_post, projwfc\_post, pp\_post, q2r\_post, matdyn\_post, pw\_phonon\_gamma, elastic\_post.}

\textbf{In-Context Example 1 (band structure with given lattice constant)}

Query: Calculate the band structure of silicon in the diamond structure (a0 = 5.43 Å).

\texttt{\textless subproblem1\textgreater}

Problem: Do an SCF calculation to converge charge density

Tool: pw\_scf

Required input: diamond Si structure

\texttt{\textless/subproblem1\textgreater}

\texttt{\textless subproblem2\textgreater}

Problem: Perform NSCF calculation along the high-symmetry path

Tool: pw\_nscf

Required input: same structure, SCF charge density

\texttt{\textless/subproblem2\textgreater}

\texttt{\textless subproblem3\textgreater}

Problem: Post-process bands to obtain band structure

Tool: bands\_post

Required input: NSCF results

\texttt{\textless/subproblem3\textgreater}

\textbf{In-Context Example 2 (equilibrium lattice constant unknown)}

Query: Calculate the equilibrium lattice constant of Na in the BCC structure.

\texttt{\textless subproblem1\textgreater}

Problem: Find equilibrium lattice constant by relaxing the cell volume and atomic positions

Tool: pw\_vc\_relax

Required input: BCC Na structure

Output: equilibrium lattice constant (Å)

\texttt{\textless/subproblem1\textgreater}

\textbf{Now handle this query:}

Query: \{question\}

- Do NOT include reasoning, explanations, or justification.

- The output must ONLY be 
\texttt{\textless subproblemN\textgreater...\textless/subproblemN\textgreater} blocks, nothing else.

\end{tcolorbox}

\tool{} will parse these \texttt{\textless subproblem\textgreater} blocks and execute each subproblem accordingly. Further extension of additional executables can be incorporated by adding corresponding options explicitly in the prompt, along with executable descriptions and parameter specifications, as described in Section~\ref{sec: parameter guess}.

\subsection{Pareto-aware Parameter Guess and Script Generation}
\label{sec: parameter guess}

For each subproblem, a parameter-guessing stage is first performed to estimate the key input parameters required for execution. By explicitly guiding the model to generate Pareto-aware parameter estimates, \tool{} achieves a principled trade-off between numerical accuracy and computational efficiency. Additionally, \tool{} aggregates the execution results of previous subproblems together with relevant internal memory, forming the \texttt{{previous\_memory}} field to enable enhanced context-aware parameter estimation and decision making. The parameter guessing prompt is designed as follows:

\begin{tcolorbox}[colback=gray!5,colframe=black!50,boxrule=0.5pt,arc=2pt]
\scriptsize

You are a computational materials scientist solving a DFT subproblem.
Now you are asked to propose ONLY the minimal necessary input parameter guesses for this subproblem.

\textbf{Instructions}

1) Propose only the parameters that are directly required or explicitly important for this query.

2) Make reasonable guesses based on existing data. Avoid performing convergence tests; instead, provide an educated estimate.

3) When guessing parameters, aim for a Pareto-efficient trade-off:
choose values that are accurate and numerically stable (i.e., likely to converge),
while keeping computational cost (time and memory) reasonably low.

\textbf{Output Schema}

\begin{verbatim}
{
  "material": "<element/compound>",
  "structure": {
    "prototype": "<fcc|bcc|rocksalt|diamond|...>"
  },
  "parameter_guesses": {
    "<only the parameters directly relevant to the query>": "<value>"
  }
}
\end{verbatim}

\textbf{The whole user query is:}

\{query\}

\{previous\_memory\}

\textbf{Current Problem to Solve:}

\{subproblem\}

Please respond strictly in the required JSON schema, without any additional explanation.

\end{tcolorbox}

Once the parameter guesses are determined, \tool{} further generates executable-specific scripts for each subproblem.
We design a dedicated set of executable requirements, provided in the \texttt{{tool\_requirements}} field, where the script format is carefully specified by domain experts. For example, for \texttt{pw.x}, we require the model to explicitly generate the standard namelists such as \texttt{\&control}, \texttt{\&system}, \texttt{\&electrons}, \texttt{\&ions}, and \texttt{\&cell}. This modular and extensible prompt integration enables \tool{} to support additional executables and functionalities with minimal changes.
\begin{tcolorbox}[colback=gray!5,colframe=black!50,boxrule=0.5pt,arc=2pt]
\scriptsize

You are a computational materials scientist.

Given the parameter JSON below, GENERATE ONLY input files in Quantum ESPRESSO.  Wrap ALL generated inputs inside ONE 
\texttt{\textless scripts\textgreater ... \textless/scripts\textgreater} block; 
each file inside a 
\texttt{\textless script\textgreater ... \textless/script\textgreater} tag.

\{tool\_requirements\}

\{previous\_memory\}

\{previous\_run\}

Only apply targeted corrections to the script based on the errors above; do not change any other guessed parameters unnecessarily.

\textbf{The whole user query is:}

\{query\}.

\textbf{Input (parameter JSON)}

\{params\_json\}

\{query\_info\}

\textbf{Current Problem to Solve:}

\{subproblem\}.

\textbf{Output Schema}

\texttt{\textless scripts\textgreater\textbackslash n
\textless script\textgreater ...one input... \textless/script\textgreater\textbackslash n
\textless script\textgreater ...next... \textless/script\textgreater\textbackslash n
...\textless/scripts\textgreater}

\textbf{Pitfall}

Do NOT regenerate or reinterpret the entire query. Only generate new or corrected scripts corresponding to the specific subproblem(s) currently being handled.  
For example, if the overall query involves multiple tasks on the same system (e.g., \texttt{vc-relax}, \texttt{scf}, and \texttt{bandgap}), generate scripts \textbf{only} for the current subproblem (e.g., \texttt{vc-relax}). Future subproblems like \texttt{scf} or \texttt{bandgap} will be generated later.

Please provide your output containing only the text wrapped in 
\texttt{\textless scripts\textgreater} and 
\texttt{\textless script\textgreater} tags.  

You MUST output raw plain text. Do NOT format or render anything. Output must be copy-paste ready for a Unix shell or input file.  

Do NOT escape or encode any characters. In particular, NEVER output ``\&amp;''.  

Use literal Quantum ESPRESSO namelists like ``\&control'', ``\&system'', ``\&electrons'', ``\&ions'', ``\&cell''.

\end{tcolorbox}

\subsection{Automatic Parallelization}

For the generated executable script, \tool{} further performs automatic parallelization by configuring hybrid MPI/OpenMP parameters for \texttt{mpirun}.
The computational scale and workload characteristics of the target program are estimated from the \texttt{{probe\_output}} field, which is obtained by running a default \texttt{mpirun} configuration for two minutes and extracting profiling information from the output log.
Based on this runtime probe, the model selects appropriate parallel parameters to balance communication overhead and computational scalability. Meanwhile, the hardware information is integrated into the \texttt{{hardware\_description}} field in natural language, enabling the model to infer physical core counts and resource constraints. The prompt is designed as follows.

\begin{tcolorbox}[colback=gray!5,colframe=black!50,boxrule=0.5pt,arc=2pt]
\scriptsize

You are an expert in Quantum ESPRESSO (QE) parallelization for HPC systems.  
Your goal is to choose the optimal Hybrid MPI/OpenMP configuration 
(\texttt{OMP\_NUM\_THREADS}, \texttt{-nk}, \texttt{-nb}, \texttt{-ntg}).

\textbf{PARAMETER RULES}

1. \textbf{Resource Limits}:

- Identify Physical Cores from the description. IGNORE logical/hyper-threads.

- Constraint: (MPI\_RANKS) * (OMP\_NUM\_THREADS) $\le$ Total Physical Cores.

- Under-subscription: It is valid to leave some cores idle to satisfy divisibility requirements.

2. \textbf{Parallel Flags}:

- \texttt{-nk} (Pools): Parallelizes over k-points.  
  Constraint: MPI\_RANKS and k-points must be divisible by nk.
  
- \texttt{-nb} (Band Groups): Parallelizes linear algebra (diagonalization).  
  Usage: Typically for systems with many electronic bands.
  
- \texttt{-ntg} (Task Groups): Parallelizes 3D FFT grids.  
  Constraint: MPI\_RANKS must be divisible by ntg.  
  Usage: Helps distribute large FFT grids across many ranks.

3. \textbf{General Optimization Strategy (Reasoning Required)}:

- Assess System Scale: Look at \texttt{nat} (number of atoms) and k-points in the input.

- For Small Systems (few atoms), MPI communication overhead often outweighs the benefit of complex parallelization logic. Excessive splitting (high \texttt{-nb}, \texttt{-ntg} or OpenMP) can degrade performance.

- For Large Systems (many atoms), calculation and memory are the bottlenecks. Advanced flags (\texttt{-ntg}, \texttt{-nb}) and Hybrid MPI/OpenMP become essential to scale.

- Decision Process: Balance the need for parallelism against the cost of communication.

\textbf{Inputs}

1) QE input script (verbatim):  
\{input\_script\}

2) Hardware resources (natural language description):  
\{hardware\_description\}

3) Probe output (from a default run):  
\{probe\_output\}

\textbf{\# Output Format (STRICT)}

Command: export OMP\_NUM\_THREADS=\textless int\textgreater; mpirun --allow-run-as-root -np \textless int\textgreater \{exec\_path\} -nk \textless int\textgreater -nb \textless int\textgreater -ntg \textless int\textgreater -in \{input\_filename\} | tee \{output\_filename\}

\end{tcolorbox}


\end{document}